\documentstyle[aps,twocolumn]{revtex}

\begin{document}
\author{I.M. Sokolov$^{1}$, J. Klafter$^{2}$ and A. Blumen$^{1}$}
\address{$^{1}$Theoretische Polymerphysik, Universit\"{a}t Freiburg,\\
Hermann-Herder-Str. 3, D-79104 Freiburg, Germany \\
$^{2}$School of Chemistry, Tel Aviv University, Tel Aviv 69978 IsraelN}
\title{Does strange kinetics imply unusual thermodynamics?\smallskip }
\date{\today}
\maketitle

\begin{abstract}
We introduce a fractional Fokker-Planck equation (FFPE) for L\'{e}vy flights
in the presence of an external field. The equation is derived within the
framework of the subordination of random processes which leads to L\'{e}vy
flights. It is shown that the coexistence of anomalous transport and a
potential displays a regular exponential relaxation towards the Boltzmann
equilibrium distribution. The properties of the L\'{e}vy-flight FFPE derived
here are compared with earlier findings for subdiffusive FFPE. The latter is
characterized by a non-exponential Mittag-Leffler relaxation to the
Boltzmann distribution. In both cases, which describe strange kinetics, the
Boltzmann equilibrium is reached and modifications of the Boltzmann
thermodynamics are not required.
\end{abstract}

\bigskip


\bigskip 

Strange kinetics \cite{SZK,Zas} which involves diffusional anomalies, both
sub- and superlinear, and nonexponential relaxations is quite wide-spread
and has been observed in a broad range of systems \cite
{SZK,BuGe,KlaShleZu,MetzBarKla,ZUB}. The ubiquity of strange kinetics rests
upon generalization of the central limit theorem due to L\'{e}vy \cite{Lévy}%
, a generalization that puts heavy-tailed distributions on a same level of
importance as the well-known Gaussian.

Anomalous diffusion in the presence or absence of an external field has been
modelled in a number of ways, including fractional Brownian motion \cite
{Mandelbrot}, generalized diffusion equations \cite{HeP}, continuous time
random walk (CTRW) models \cite{KBS}, Langevin and generalized Langevin
equations \cite{FogQ} and generalized thermostatics \cite{Tsallis}. In
particular CTRW has been demonstrated to be a powerful approach in
describing subdiffusive as well as superdiffusive processes and in
interpreting experimental results. It is not straightforward, however, to
incorporate force fields and boundary conditions in this formalism.

An alternative approach to processes which display strange kinetics is based
on fractional equations, which are suitable for handling external fields and
for considering boundary value problems. In the case of subdiffusion it was
realized that the replacement of the local time derivative in the diffusion
equation by a fractional operator accounts for memory effects responsible
for anomalous behavior \cite{MetzBarKla,BarMetzKla}. In the presence of an
external field a fractional Fokker-Planck equation FFPE has been introduced 
\cite{MetzBarKla,BarMetzKla}:

\begin{equation}
\frac{\partial }{\partial t}P(x,t)=K_{(\alpha )}\,_{0}D_{t}^{1-\alpha }{\cal %
L}_{FP}P(x,t)  \label{subdiff}
\end{equation}
where ${\cal L}_{FP}$ is the Fokker-Planck operator, 
\begin{equation}
{\cal L}_{FP}=\frac{{\partial }^{2}}{{\partial }x^{2}}-\frac{\partial }{%
\partial x}\frac{f(x)}{k_{B}T}.  \label{FPO}
\end{equation}
$_{0}D_{t}^{1-\alpha }$ is a fractional Riemann-Liouville operator $0<\alpha
<1$, and $K_{(\alpha )}$ is a generalized \mbox{(sub-)diffusion}
coefficient, having the dimension $\left[ K_{(\alpha )}\right] =\left[ {\rm L%
}^{2}/{\rm t}^{\alpha }\right] .$ The force $f(x)$ is related to the
external potential $U(x)$ through $f(x)=-dU/dx$, and $k_{B}$ is the
Boltzmann constant. The differential operator $_{0}D_{t}^{1-\alpha }$ acting
on functions of time is defined through \cite{MeK} 
\begin{equation}
_{0}D_{t}^{1-\alpha }Z(t)=\frac{1}{\Gamma (\alpha )}\frac{\partial }{%
\partial t}\int_{0}^{t}dt^{\prime }\frac{Z(t^{\prime })}{(t-t^{\prime
})^{1-\alpha }}.  \label{RiLi}
\end{equation}
The FFPE, Eq.(\ref{subdiff}), has been derived using a Kramers-Moyal
expansion of the CTRW nonlocal equation \cite{BarMetzKla}. The solution of
this FFPE is characterized by a subdiffusive behavior and by a
nonexponential Mittag-Leffler decay of the single modes. The decoupled
structure of Eq.(\ref{subdiff}) guarantees that the Boltzmann distribution
is attained at equilibrium \cite{MetzBarKla,BarMetzKla,MeK}. We note that
the latter is also a property of the regular Fokker-Planck equation
corresponding to $\alpha =1$.

Less clear has been the situation for FFPEs which correspond to L\'{e}vy
spatial {\it flights}. Previously proposed equations \cite{Zas,FogQ} seem
not to lead to the Boltzmann distribution, a point whose impact has been
overlooked. This might suggest therefore that strange kinetics requires
unusual thermodynamics \cite{Tsallis}. Here we derive a FFPE for L\'{e}vy
flights in the presence of an external force. Our starting point is a
representation of L\'{e}vy flights in terms of a subordination of random
processes \cite{Feller,S2000}. This representation corresponds to processes
in which space and time are decoupled and it does not account for L\'{e}vy
walks \cite{SZK,KlaShleZu,KBS}. Namely, in what follows we obtain a
diverging mean-square displacement in the force-free case. The solution of
the FFPE which we derive again leads to the Boltzmann distribution in the
equilibrium limit, re-emphasizing that there is no need to modify
conventional thermodynamics in order to obtain strange kinetics. We bring
some examples for solving this FFPE for boundary value problems.

As we proceed to show, the corresponding generalization of the Fokker-Planck
equation for L\'{e}vy flights is: 
\begin{equation}
\frac{\partial }{\partial t}P(x,t)=-K^{(a)}\left( -{\cal L}_{FP}\right)
^{\alpha }P(x,t),  \label{super}
\end{equation}
where the operator $\left( -{\cal L}_{FP}\right) ^{\alpha }$ is the $\alpha $%
-th power of the operator $-{\cal L}_{FP}=-\partial ^{2}/\partial
x^{2}+\partial /\partial x(f(x)/k_{B}T)$ as will be derived below, and the
corresponding generalized \mbox{(super-)diffusion} coefficient has as
dimension $\left[ K^{_{(\alpha )}}\right] =\left[ {\rm L}^{2\alpha }/{\rm t}%
\right] $.

The CTRWs can be viewed as Markovian random walks on a lattice (with lattice
constant $a$) given in terms of the number of steps $n$ of the random
walker. $P(x,n)$ is a probability distribution function (pdf) of the
particles' displacement $x$ after $n$ steps. The number of steps $n$
performed during the time $t$ follows the probability distribution $S(n,t)$,
which may include memory effects \cite{BKZ}. The overall displacement during
time $t$ is then given by 
\begin{equation}
P(x,t)=\sum_{n=0}^{\infty }P(x,n)S(n,t).  \label{Subord}
\end{equation}
In the force-free case the pdf $P(x,n)$ corresponds typically to normal
diffusion behavior, and thus $\overline{x^{2}}\propto n$. On the other hand,
the typical number of steps can grow sub- or superlinearly in time, so that
the overall behavior can be anomalous.

Here we concentrate on the superdiffusive case and assume that the random
process $\left\{ n(t)\right\} $ is characterized by a diverging mean density
of events, so that the first moment of the number $n$ of steps does not
exist. As a realization of such a process we can take that the numbers of
jumps during different time intervals of unit length are independent random
variables distributed according to $S(n,1)\propto n^{-1-\alpha }$. For $t$
large enough the distribution $S(n,t)$ tends then to a stable L\'{e}vy-law $%
L(n;\alpha ,\beta )$ \cite{Feller}. Since $n$ is nonnegative, this law is
the one-sided extreme distribution for which $\beta =-\alpha $ ($0<\alpha <1$%
). If different time intervals $t$ are considered, the distribution $S(n,t)$
scales as 
\begin{equation}
S(n,t)=\frac{1}{t^{1/\alpha }}L\left( \frac{n}{t^{1/\alpha }};\alpha
,-\alpha \right) .
\end{equation}
Imagine now a random walker moving under the influence of a weak force $f(x)$%
. Such a force introduces an asymmetry into the walker's motion, since the
probabilities for forward and backward jumps, $w_{+}$ and $w_{-}$ are now
weighed with the corresponding Boltzmann-factors, $w_{+}/w_{-}=\exp
(fa/k_{B}T)$. For small $f$ one can take $w_{+}=1/2+fa/2k_{B}T$ and $%
w_{-}=1/2-fa/2k_{B}T$. Note that the process described in such a way is a
Markovian one, and can be characterized by a transition probability 
\begin{equation}
W(x,t+\Delta t\left| x^{\prime },t\right. )=\sum_{n=0}^{\infty
}P(x-x^{\prime },n)S(n,\Delta t).
\end{equation}
For $\Delta t$ in the intermediate range, i.e. large enough to view both $x$
and $n$ as being continuous and to approximate $P(x,n)$ by the Gaussian $%
P(x,n)=\left( 2\pi n\right) ^{-1/2}\exp \left( -(x-vn)^{2}/2a^{2}n\right) $
with $v=fa/2k_{B}T$, yet small enough to have the typical displacement small
on the scale of change of $f(x)$, one obtains: 
\begin{eqnarray}
&&W(x,t+\Delta t\left| x^{\prime },t\right. )=  \nonumber \\
&=&\int_{0}^{\infty }\frac{1}{\sqrt{2\pi n}}\exp \left( -\frac{(x-x^{\prime
}-vn)^{2}}{2a^{2}n}\right) S(n,\Delta t)dn.
\end{eqnarray}
The overall Markovian process is then governed by the integral
Chapman-Kolmogorov equation 
\begin{equation}
P(x,t+\Delta t)=\int W(x,t+\Delta t\left| x^{\prime },t\right. )P(x^{\prime
},t)dx^{\prime }.  \label{ChaK}
\end{equation}
Let us concentrate on the long-time, large $x$, behavior of the system and
take the force $f$ to be smooth. On such scales we consider the $x$%
-Fourier-transform of Eq.(\ref{ChaK}) and get: 
\begin{eqnarray}
&&P(k,t+\Delta t)=  \label{exp} \\
&=&\int_{0}^{\infty }dn\exp \left( -(ikv+k^{2})a^{2}n\right) S(n,\Delta
t)P(k^{\prime },t).
\end{eqnarray}
For random processes leading to diffusive behavior, the first moment of the
distribution $S(n,\Delta t)$ for small $\Delta t$ exists, so that one can
expand for small $k$ the exponential into a power series, getting: 
\begin{eqnarray}
P(k,t+\Delta t) &=&\int_{0}^{\infty }dn\left( 1-(ikv+k^{2})a^{2}n\right)
S(n,\Delta t)P(k,t)  \nonumber \\
&=&\int_{0}^{\infty }dn\left( 1-(ikv+k^{2})a^{2}\left\langle n\right\rangle
_{\Delta t}\right) P(k,t)  \nonumber \\
&=&P(k,t)-(ikv+k^{2})a^{2}\left\langle n\right\rangle _{\Delta t}P(k,t).
\label{expand}
\end{eqnarray}
(a Kramers-Moyal procedure). For normal diffusive processes one has $%
\left\langle n\right\rangle _{\Delta t}\simeq w\Delta t$, where $w$ is the
jumping rate, so that in the continuum limit 
\begin{equation}
\frac{\partial }{\partial t}P(k,t)=-K\left( ik\frac{f}{k_{B}T}+k^{2}\right)
P(k,t),  \label{Fourier1}
\end{equation}
with $K=a^{2}w/2$ being the diffusion coefficient. In the $x$-representation
this is the conventional Fokker-Planck equation (FPE) \cite{Risken}: 
\begin{equation}
\frac{\partial P(x,t)}{\partial t}=K\left( -\frac{\partial }{\partial x}%
\frac{f}{k_{B}T}+\frac{\partial ^{2}}{\partial x^{2}}\right) P(x,t).
\end{equation}
In the case when $S(n,\Delta t)$ is a L\'{e}vy-stable law of index $\alpha $%
, $0<\alpha <1$, the first moment of $n$ diverges, and the series expansion
of the exponential, Eq.(\ref{expand}), is not possible. On the other hand
for $\alpha <1$ the integral $\phi (k)=\int_{0}^{\infty }\exp (-\kappa \tau
)S(\tau ,\Delta t)d\tau $ converges for each $\kappa =\xi +i\eta $, $%
\mbox{Re}\xi >0$, and is a stretched-exponential function \cite{Feller}. For
extreme L\'{e}vy-stable distributions with $0<\alpha <1$ (those which
vanishing identically for negative arguments) one has $\phi (k)=\exp
(-\kappa ^{\alpha })$. Thus, performing the integration in Eq.(\ref{exp})
one gets: 
\begin{equation}
P(k,t+\Delta t)=\exp \left[ -K_{(\alpha )}\left( ik\frac{f}{k_{B}T}%
+k^{2}\right) ^{\alpha }\Delta t\right] P(k^{\prime },t)
\end{equation}
Expanding now the exponential and repeating the steps leading to Eq.(\ref
{expand}) we have 
\begin{equation}
\frac{\partial P(k,t)}{\partial t}=-K_{(\alpha )}\left( ik\frac{f}{k_{B}T}%
+k^{2}\right) ^{\alpha }P(k,t)  \label{Fourier}
\end{equation}
Comparing the terms $-(ikf/k_{B}T+k^{2})^{\alpha }$ and $-(ikf/k_{B}T+k^{2})$
in Eqs. (\ref{Fourier}) and (\ref{Fourier1}), which represent the
corresponding transport operators ${\cal L}_{\alpha }$ and ${\cal L}_{FP}=%
{\cal L}_{1}$ in Fourier space, we see that they are connected by the
relation ${\cal L}_{\alpha }=-\left( -{\cal L}_{FP}\right) ^{\alpha }$. The
same relation holds, of course, when one shifts to the $x$-representation:

\begin{equation}
\frac{\partial P(x,t)}{\partial t}=-K_{(\alpha )}\left( -{\cal L}%
_{FP}\right) ^{\alpha }P(x,t),  \label{FFPE}
\end{equation}
see note \cite{Note}. Note that Eq.(\ref{FFPE}) differs from the expressions
proposed in \cite{FogQ} where either only the second part of the
Fokker-Planck operator (a $\Delta $-term) is changed (and corresponds in our
notation to $-(-\partial ^{2}/\partial x^{2})^{\alpha }$) or where a sum of
two terms is introduced, so that fractional space derivatives of the orders $%
\alpha $ and $2\alpha $ appear. Note that in general ${\cal L}_{\alpha }\,$%
can not be decoupled into additive parts responsible separately for drift
and for diffusion.

Some important properties of the L\'{e}vy-diffusion in the presence of a
force field stem from Eq.(\ref{FFPE}). Since $-{\cal L}_{FP}$ and ${\cal L}%
_{\alpha }$ commute with each of their powers, the eigenfunctions of these
operators coincide. The corresponding eigenvalues of ${\cal L}_{\alpha }$
are those of $-{\cal L}_{FP}$ raised to the power of $\alpha $: 
\begin{equation}
\lambda _{k}^{FFP}=-(-\lambda _{k}^{FP})^{\alpha }.  \label{EVs}
\end{equation}
Note that the eigenfunctions of $-(ikf/k_{B}T+k^{2})$ and of $%
-(ikf/k_{B}T+k^{2})^{\alpha }$ (describing conventional FPE and an FFPE in
an infinite homogeneous system respectively) can be chosen to be the same.
Exemplarily, $\exp (ikx)$ is the eigenfunction of free motion in both cases;
we denote its eigenvalues by $\lambda _{k}^{FP}$ and $\lambda _{k}^{FFP}$
respectively. Thus, if ${\cal L}_{FP}$ has a (nondegenerate) zero
eigenvalue, whose eigenfunction corresponds to a stationary state, the same
holds for ${\cal L}_{\alpha }$. The stationary states of the systems
described by FPE and by FFPE therefore coincide. For closed systems (no
currents at infinity), the stationary state is that of thermodynamic
equilibrium and is given by the Boltzmann distribution. This is a general
property of each subordination process, since a state stationary in $t$ is
also stationary in $n$.

The solution of FFPEs under the given initial and boundary conditions can be
obtained by means of an eigenfunction expansion, as is generally the case
for normal and subdiffusive motion \cite{MetzBarKla,BarMetzKla,MeK,Risken}.
If $\phi _{m}(x)$ are the eigenfunctions of the Fokker-Planck operator, then
the solution of FFPE can be expressed as 
\begin{equation}
P(x,t)=\sum_{m}a_{m}\phi _{m}(x)\Phi _{m}(t)
\end{equation}
where $\Phi _{m}(t)$ are the corresponding temporal decay forms. Here the
difference between sub- and superdiffusive FFPE gets to be evident: in the
subdiffusive case $\Phi _{m}(t)$ are solutions of a fractional ordinary
differential equation 
\begin{equation}
\frac{d}{dt}\Phi _{m}(t)=K\lambda _{m}\,_{0}D_{t}^{1-\alpha }\Phi _{m}(t)
\end{equation}
($\lambda _{n}$ being real and negative). Hence the $\Phi _{m}(t)$ are
Mittag-Leffler functions \cite{Tsallis,BarMetzKla,MeK}. On the other hand
the superdiffusive FFPE (being of first order in time) leads to 
\begin{equation}
\frac{d}{dt}\Phi _{m}(t)=-K_{(\alpha )}\left( -\lambda _{m}\right) ^{\alpha
}\,\Phi _{m}(t)  \label{temporal}
\end{equation}
corresponding to a simple exponential relaxation, $\Phi _{m}(t)=\exp
(-K_{(\alpha )}\left| \lambda _{m}\right| ^{\alpha }t)$. Thus, in the case
of a {\it discrete spectrum} and of real, negative $\lambda _{m}$ the
L\'{e}vy-flight FFPE retains the exponential nature of the relaxation to
equilibrium, a behavior typical for normal FPE, so that only the
corresponding relaxation times change. For example, the relaxation behavior
of a particle in a harmonic potential, $f(x)=-\gamma x$, follows immediately
from a standard solution of the FPE \cite{Risken}: The eigenfunctions can be
expressed through those of the Schroedinger equation, and the spectrum
consists of a zero eigenvalue, $\lambda _{0}=0$, and of equidistant negative
eigenvalues, $\lambda _{n}=-(\gamma /k_{B}T)n$. Since the spectrum of a
Fokker-Planck operator with a harmonic potential is discrete, the relaxation
is multiexponential. The equilibrium state of such a system (the
eigenfunction corresponding to $\lambda _{0}=0$) shows a Boltzmann
distribution. The longest relaxation time is given by the first eigenvalue, $%
\lambda _{1}=-\gamma /k_{B}T$, so that $\tau =(k_{B}T/\gamma )^{\alpha
}/K_{(\alpha )}$.

Another interesting example corresponds to the motion in the absence of a
field of a particle in an interval with absorbing boundaries at $x=\pm l$.
The eigenfunctions of the Fokker--Planck operator are now the trigonometric
functions, $\phi _{m}(x)=\cos \left[ (m+1/2)\pi x/l\right] $, and the
corresponding eigenvalues are $\lambda _{m}=-\left[ (m+1/2)\pi /l\right] ^{2}
$. The eigenvalues of ${\cal L}_{\alpha }$ are $\lambda _{m}=-K_{(\alpha
)}\left[ (m+1/2)\pi /l\right] ^{2\alpha }$, so that the overall relaxation
again follows a multiexponential pattern. The survival probability for a
particle initially situated at the middle of the interval, $x=0$, is equal
to 
\begin{eqnarray}
P(t) &=&\sum_{m=0}^{\infty }\int_{-l}^{l}\cos \left[ (m+1/2)\frac{\pi }{l}%
x\right] e^{\lambda _{m}t}= \\
&=&\sum_{n=0}^{\infty }\frac{4}{\pi }\frac{(-1)^{n}}{(2m+1)}e^{-K_{\alpha
}\left[ (n+1/2)\pi /l\right] ^{2\alpha }t}.
\end{eqnarray}
At longer times this decay tends to a simple exponential with the
characteristic time $\tau =K_{(\alpha )}^{-1}(l/2\pi )^{2\alpha }$. Note
that the $l$-dependence of this characteristic time differs from that
encountered in normal diffusion, where $\tau =K^{-1}(l/2\pi )^{2}$. In the
case $\alpha =1/2$ a simple analytical expression holds at all times: $%
P(t)=\arctan \left[ \exp \left( -\frac{\pi }{2}\frac{K_{(1/2)}}{l}t\right)
\right] $, see Eq. 5.2.4.8 of Ref.\cite{BP}.

Using a representation of L\'{e}vy flights in terms of a subordination of
random processes and following the Kramers-Moyal procedure we have derived a
fractional Fokker-Planck equation for L\'{e}vy flights. It has been shown
that when the regular Fokker-Planck operator has a discrete spectrum (as
happens under appropriate potentials or boundary conditions) anomalous
transport results in an exponential relaxation towards an equilibrium
distribution. These properties of the L\'{e}vy-flight FFPE are compared with
earlier findings for subdiffusive FFPE. The latter is characterized by a
non-exponential Mittag-Leffler relaxation. The equilibrium solution
corresponds in both cases to the Boltzmann distribution, emphasizing that
there is no need to modify conventional thermodynamics in order to obtain
strange kinetics.

The authors gratefully acknowledge the support of the German-Israeli
foundation (GIF), of the DFG through SFB428 and of the Fonds der Chemischen
Industrie.

\end{document}